# Helical magnetism and structural anomalies in triangular lattice α-SrCr$_2$O$_4$

S E Dutton,[1,*] E Climent-Pascual,[1] P W Stephens,[2] J P Hodges,[3] A Huq,[3] C L Broholm[4] and R J Cava.[1]

[1] Department of Chemistry, Princeton University, Princeton, New Jersey 08544, USA

[2] Department of Physics and Astronomy, Stony Brook University, Stony Brook, New York 11794, USA

[3] Neutron Scattering Science Division, Oak Ridge National Laboratory, P.O. Box 2008, MS-6475, Oak Ridge, Tennessee 37831, USA

[4] Department of Physics and Astronomy, The Johns Hopkins University, Bloomberg 345, 3400 North Charles Street, Baltimore, Maryland 21218, USA

[*] Corresponding author

Email: sdutton@princeton.edu

**Abstract.** α-SrCr$_2$O$_4$ has a triangular planar lattice of $d^3$ Cr$^{3+}$ made from edge sharing CrO$_6$ octahedra; the plane shows a very small orthorhombic distortion from hexagonal symmetry. With a Weiss temperature of –596 K and a three-dimensional magnetic ordering temperature of 43 K, the magnetic system is quasi two-dimensional and frustrated. Neutron powder diffraction shows that the ordered state is an incommensurate helical magnet, with an in-plane propagation vector of **k** = (0,0.3217(8),0). Temperature dependent synchrotron powder diffraction characterization of the structure shows an increase in the inter-plane spacing on cooling below 100 K and an inflection in the cell parameters at the magnetic ordering temperature. These anomalies indicate the presence of a moderate degree of magneto-structural coupling.

# 1. Introduction

Interest in the low temperature structural and magnetic behaviour of materials with triangular lattices arises from the intrinsic geometric magnetic frustration[1] and the resultant delicate balance between competing magnetic phases that can result in novel magnetic order[2-5] or the absence of long-range ordering.[6] Unlike the kagome lattice, a two-dimensional triangular lattice of classical spins has a unique lowest energy state, with the spins arranged at 120° to one another.[7] However the large number of states close in energy to the ground state allow quantum effects and weaker sub-leading interactions to produce a range of different types of magnetism on triangular lattices.[8, 9, 4, 5, 10, 11] Distortions of the ideal triangular lattice can give rise to incommensurate magnetic ordering with a spiral or helical magnetic ground state, as is observed in $CuFeO_2$ delafossite[10, 11] and $KFe(MoO_4)_2$.[4] The exact behaviour of these systems is highly sensitive to the structure of the materials. For example, changing the size of the inter-layer separation in $AFe(MoO_4)_2$ compounds perturbs the magnetism significantly; $RbFe(MoO_4)_2$ has the 120° ordering expected for a triangular lattice,[12, 13] while $KFe(MoO_4)_2$ forms two distinct magnetic networks.[4, 5]

The structure of $\alpha$-$SrCr_2O_4$ was previously reported,[14] but its magnetic properties have not been characterized. This material has a layered structure with triangular sheets of formula $CrO_2$ separated by $Sr^{2+}$ in trigonal prismatic co-ordination (Figure 1). At room temperature there is a small distortion in the triangular $Cr^{3+}$ lattice, which occurs as an indirect effect of the rectangular ordering of the $Sr^{2+}$ between the layers; the result is overall orthorhombic symmetry. The magnetic properties of the analogous Ca compound $\alpha$-$CaCr_2O_4$ have recently been reported.[8] In that compound the $Cr^{3+}$ moments form an antiferromagnetic ground state with helical magnetic ordering at 43 K. $\alpha$-$SrCr_2O_4$ and $\alpha$-$CaCr_2O_4$ are part of a family of isostructural compounds that also includes a Ba variant.[15, 14, 16] The larger alkaline earth versions have an expanded crystallographic unit cell, primarily manifesting as an increase in the separation of the $CrO_2$ layers. As has been reported for $ACrO_2$ ($A$ = Li, Na and K[17]), this change in inter-planar spacing may be expected to result in changes in the magnetic properties. Here we report the characterization of the low temperature behaviour of $\alpha$-$SrCr_2O_4$. As for the Ca analogue, magnetic ordering in $\alpha$-$SrCr_2O_4$ is observed on cooling below 43 K, with an incommensurate propagation vector, **k** = (0,0.3217(8),0) consistent with helical magnetic ordering similar to that of $\alpha$-$CaCr_2O_4$. Thus the increase in the inter-planar spacing in this system on going from Ca to Sr has minimal effect on the magnetic ordering. With the onset of three-dimensional magnetic ordering, an inflection in the temperature dependence of all three lattice parameters is observed. Further, the perpendicular-to-plane lattice parameter of $\alpha$-$SrCr_2O_4$ shows an anomalous continuous increase on cooling below 100 K, which reflects a slight increase in the puckering of the $CrO_2$ layers. These structural changes, though they appear to occur without changing the average structure, may help to relieve the magnetic frustration through local magnetostructural coupling.

## 2. Experimental

Powder samples of α-SrCr$_2$O$_4$ were prepared using a ceramic synthesis route. A stoichiometric mixture of SrCO$_3$ (Sigma Aldrich, 99.9+%) and Cr$_2$O$_3$ (Alfa Aesar, 99.97%) was intimately mixed and then pressed into pellets. The pellets were heated at 50 °C h$^{-1}$ to 1000 °C under a dynamic vacuum of < 10$^{-5}$ Torr. After dwelling at 1000 °C for 2 hours, the furnace was turned off and when at room temperature the chamber was filled with Ar. The sample was then heated to 1550 °C for 12 hours. Initial characterisation using a Bruker D8 Focus diffractometer operating with Cu Kα radiation and a graphite diffracted beam monochromator indicated the formation of phase pure α-SrCr$_2$O$_4$.

Magnetic susceptibility and specific heat measurements were made using a Quantum Design Physical Properties Measurement System (PPMS). Field dependent magnetization (M) measurements at temperatures between 2 and 300 K showed linear behaviour up to μ$_0$H = 9 T, and therefore a μ$_0$H = 5 T field was employed for measurement of the magnetic susceptibility, χ, defined as M/H. Temperature dependent magnetization measurements of finely ground α-SrCr$_2$O$_4$ were collected between 2 and 300 K after cooling in zero field. Specific heat was measured in zero field from $2 \leq T \leq 300$ K. The pellet for heat capacity was prepared by heating for 12 hours at 1550 °C under a static argon atmosphere in a vacuum furnace.

The temperature dependence of the lattice parameters was measured by synchrotron powder X-ray diffraction (SXRD) at beam line X16C at the National Synchrotron Light Source at Brookhaven National Laboratory. A sample was loaded in a glass capillary, $\phi$ = 0.3 mm, and diffraction patterns (λ ~ 0.70 Å) were collected between 20 and 280 K. To account for possible wavelength drift during the measurements, an internal silicon standard was employed. Rietveld refinement[18] of the structure was carried out using the Topaz-Academic suite of programs.[19] Backgrounds were fitted using a Chebyshev polynomial of the first kind and the peak shape was modeled using a pseudo-Voigt function.

Neutron powder diffraction measurements were carried out at the Spallation Neutron Source (SNS) at Oak Ridge National Laboratory on the POWGEN time of flight diffractometer. The sample was loaded in a vanadium can, $\phi$ = 8 mm, and patterns were collected with wavelength centered at 2.132 Å at 100 K, 50 K and 12 K. To explore the evolution of the magnetic peaks as a function of temperature, shorter scans were also collected at several temperatures below $T_N$. Rietveld refinement was carried out using the Fullprof program.[20] The background was fit using a six coefficient polynomial function and the peak shape was modeled based on a convolution of back to back exponentials with a pseudo-Voigt function with d-space dependencies of the peak shape and positions appropriate for a cold cryogenic moderator. Additional terms to model anisotropic peak broadening as described by Stephens[21] were included in the refinement; this contribution was fixed to be purely Lorentzian, and all of the six terms allowed by the orthorhombic symmetry P*mmn* unit cell were refined.

## 3. Results

Magnetic susceptibility measurements on α-SrCr$_2$O$_4$ show an antiferromagnetic-like transition with $T_N \sim 43$ K (Figure 2). Above $T_N$ the magnetic susceptibility obeys the Curie-Weiss law, $\chi - \chi_o = C/(T-\theta)$, and fits were carried out for T > 150 K. Reflecting the two-dimensional and geometrically frustrating triangular arrangement of the Cr$^{3+}$, the Weiss Temperature, $\theta = -596$ K, is large relative to $T_N$ (f = $\theta/|T_N|$ = 13.9). The obtained Curie constant, $C$ = 2.01 emu Oe$^{-1}$ mol$_{Cr}^{-1}$ K$^{-1}$ and effective magnetic moment, $\mu_{eff}$ = 4.01 $\mu_B$, are slightly higher than might be expected for paramagnetic Cr$^{3+}$ ($C$ = 1.875 emu Oe$^{-1}$ mol$_{Cr}^{-1}$ K$^{-1}$, $\mu_{eff}$ = 3.87 $\mu_B$). This can be attributed to the relatively small temperature range over which fitting to the Curie-Weiss law was carried out and the presence of long-range magnetic fluctuations when $T < |\theta|$. The specific heat of α-SrCr$_2$O$_4$ is shown in Figure 3. A sharp feature is observed at $T \sim T_N$; no features are observed at higher temperatures. At low temperatures, $T$ < 20 K, the specific heat can be modeled using a power law, $C_p = aT^b$, with $b$ = 2.98(2), indicating that three-dimensional magnetic ordering occurs below $T_N$. The absence of non-magnetic structural analogues of α-SrCr$_2$O$_4$ does not allow for a direct deduction of the lattice contribution to the specific heat and so instead the lattice contribution at low temperature, $T$ < 75 K was approximated using a Debye model. The calculated change in entropy is shown as an inset in Figure 3. At $T_N$, the inferred magnetic entropy increases much less than the full magnetic entropy associated with S = 3/2 Cr$^{3+}$ (2Rln(2S+1) ~ 23 J K$^{-1}$ mol$^{-1}$), indicating the persistence of two-dimensional or short-range ordering above $T_N$.

The temperature dependence of the lattice parameters of α-SrCr$_2$O$_4$ is shown in Figure 4. Throughout the measured temperature range, the $c/b$ ratio remains constant (~ 0.87) and close to the value expected for an ideal hexagonal lattice (for an ideal hexagonal sheet $c/b$=cos30 = 0.87). Thus the geometry of the CrO$_2$ sheets is close to that of an ideal hexagonal layer. At temperatures between 300 and 100 K, the expected decrease in the unit cell parameters is observed on cooling. At $T \sim 100$ K the $a$ parameter, which is the measure of twice the inter-layer spacing, shows an anomalous increase; no anomalous features are observed in the $b$ and $c$ parameters, which characterize the layers. The increase in the inter-layer spacing is however only ~ 0.01 Å, or 0.02%, on cooling from 100 to 12 K, this is much smaller than the in-plane contraction of ~ 0.12% over the same temperature range, hence the unit cell volume continues to decrease and no negative volume thermal expansion is observed. At $T \sim T_N$ an inflection point in the temperature dependence of all three lattice parameters is observed. In spite of these anomalies in the thermal expansion of α-SrCr$_2$O$_4$, no changes in the symmetry are observed in either the X-ray or neutron diffraction data, with the P$mmn$ unit cell previously described by Pausch and Müller-Buschbaum[14] consistent with the diffraction data from 12 K to 300 K.

A fit to the neutron diffraction data collected at 50 K is shown in Figure 5. Structural parameters and bond lengths from refinement of patterns collected at 12, 50 and 100 K are given in Tables 1 and 2.

Consideration of the connectivity of the Cr lattice shows a triangular lattice with only a minor distortion. The distortions are reflected in the displacements of the refined Cr1 position from an ideal hexagonal pseudosymmetry site at (½,¼,½) and the non-ideal ratio of the in-plane cell parameters. Within the $CrO_2$ layers, the orthorhombic symmetry allows the Cr1 to move in the $c$ direction while the $y$ co-ordinate is fixed. Further, the site symmetry of Cr1 permits it to move out of the $x = ½$ plane, at all temperatures this is observed as a slight puckering of the the $CrO_2$ sheets which increases on cooling. The net result is that both the in-plane and inter-plane Cr-Cr distances are very slightly different (see Table 2). Over the measured temperature range, an inversion of the short and long interactions within individual triangles in the plane occurs. When averaged over the whole sheet these changes do not reflect any significant variation in the structure of the $CrO_2$ layers.

A broadening of some of the peaks in the neutron diffraction patterns is observed, which is modeled by the refinement of anisotropic strain parameters. These parameters show a small value (< 0.002 a.u.) associated with $h00$ reflections, intermediate values (0.1-0.2 a.u.) for $0k0$, $hk0$, $h0l$ reflections and larger values (0.4-0.6 a.u.) associated with $00l$ and $0kl$ reflections. The strain thus appears to be primarily associated with the $c$ lattice direction. Comparison of the values obtained for α-$SrCr_2O_4$ with those obtained from other triangular lattices[22, 23] shows that the anisotropic broadening is relatively small. Almost constant values of the strain parameters are obtained in refinement of the neutron powder diffraction data collected between 12 and 100 K; only the $S_{202}$ and $S_{220}$ parameters show a significant change, decreasing and increasing on cooling respectively. These two values reflect the coupling between one of two directions within the $CrO_2$ sheets with the perpendicular $x$-axis and may reflect how the unusual temperature dependence of the $a$ parameter affects the geometry of the $CrO_2$ layers. Since the strain is primarily associated with the in-plane reflections it is unlikely that stacking faults are the origin of the strain. Strain of this kind has previously been associated with quench strain from a higher temperature structural phase transition.[24] Alternatively the apparent strain may arise from small domains in α-$SrCr_2O_4$ with monoclinic or triclinic symmetry which when spatially averaged yield orthorhombic symmetry, such as found in $V_{1-x}Mo_xO_2$ below its metal to insulator transition.[25] Given the strain associated with the orthorhombic structure at low temperatures, it may be that a structural phase transition to an ideal orthorhombic phase occurs above room temperature. At still higher temperatures, the Sr may disorder and a hexagonal symmetry phase may occur. While this has not been reported for any of the $ACr_2O_4$ materials,[8, 15, 14, 16] structurally related β-$SrRh_2O_4$ crystallizes in the hexagonal $P\bar{6}2c$ space group,[26] which has the Sr disordered over two sites between the $CrO_2$ sheets.

On cooling below $T_N$ additional broad peaks are observed in the neutron diffraction pattern, the two most intense close to d ~ 4.15 Å (Figure 6a). The magnetic scattering can be approximately indexed using the propagation vector, $\mathbf{k} \sim (0,⅓,0)$ and described using the same magnetic model as proposed for α-$CaCr_2O_4$. In this model (Figure 6b), the spins are constrained within the $ac$-plane and form a spiral arrangement that propagates along the *b*-direction. Within the layers, the residual moment from each individual triangle is zero when **k** = (0,⅓,0). Adjacent layers are coupled antiferromagnetically. This magnetic model gives a reasonable fit to the data, however the two peaks observed close to d ~ 4.15 Å are poorly modeled (Figure 6a inset). An incommensurate magnetic ordering, with propagation vector **k** ~ (0,0.3217(8),0) was found to model the data much better. On warming from low temperatures towards $T_N$, the intensity of the magnetic reflections decreases (Figure 6c) and just below $T_N$, at 40 K, a magnetic moment of 1.75(4) $\mu_B$ is observed; a significant reduction from the 2.34(3) $\mu_B$ observed at 12 K. At 50 K no residual intensity from the magnetic phase is observed, and at d ~ 4.15 Å the intensity cannot be distinguished from the background. The broadening of the magnetic Bragg peaks was originally modeled using the same model for anisotropic broadening as in the structural phase however this did not fully account for the peak shape. The microstrain was thus constrained to be the same as for the nuclear phase and the additional broadening modeled using a spherical harmonic expansion consistent with Laue class *mmm*. The broadening of the peaks in the magnetic phase is thus an indication of a finite magnetic correlation length. No reduction in the width of the magnetic reflections is observed on cooling and thus the magnetic correlation length in the ordered phase remains constant on cooling from 40 K to 12 K, which is 31 K below $T_N$. Reduced magnetic correlation lengths are commonly observed in two-dimensional magnets as a consequence of weak disorder and often occur despite the specific heat indicating long-range magnetic ordering.[8, 3] Due to the anisotropic strain present in both the magnetic and structural phases it is difficult to approximate the magnetic correlation length by modeling the peaks, especially given the degree of overlap between the two most intense magnetic reflections. However a rough approximation was made by fitting three reflections, two magnetic and one structural, close to d~4.15 Å using a lorentzian peak shape. By constraining the width of the two magnetic reflections to be the same and accounting for instrumental broadening using the peak from the structural phase a magnetic correlation length of $\xi$=400(200) Å was obtained.

**4. Discussion**

Expansion of the lattice parameter describing the inter-layer spacing on cooling is not unique to α-SrCr$_2$O$_4$; in the related delafossite family, Cu*M*O$_2$ (*M* = Cr, Al, In, La, and Sc) negative thermal expansion is observed.[27-29] The delafossites are structurally very similar to α-SrCr$_2$O$_4$, being comprised of sheets of *M*O$_2$ interspaced by linearly coordinated Cu$^+$. In CuCrO$_2$ it is a distortion of the CrO$_6$ octahedra on cooling which causes the expansion of the inter-layer spacing.[27] In the Cu*M*O$_2$ (*M* = Al, In, La, and Sc) compounds[28, 29] however, the expansion on cooling is attributed to a softening of a phonon mode which increases the Cu-O bond length. From our low temperature neutron diffraction measurements there is no indication of an increase in the size of the SrO$_6$ trigonal prism. However a small change in the geometry of the Cr1 octahedra is seen and as the Cr1-O1 bond length increases a concomitant decrease in the Cr1-O2 interaction is observed. This distortion is associated with an increase in the displacement of the Cr1 position from the ideal (½,¼,½) hexagonal position on cooling. The distortion also induces an increase in the difference between the two inter-layer Cr1-Cr1 interactions, puckering the $CrO_2$ sheets. As a result of the distortion there is an increase in the disparity between the inequivalent Cr1-Cr2 interactions, changing the balance of the magnetic interactions within the $Cr^{3+}$ lattice. For isotropic spin $Cr^{3+}$ ($d^3$) cations, structural distortions due to orbital ordering are not expected and as such this distortion may be driven by the tendency to singlet formation. A spin-Peierls distortion is associated with the structural transition in the $ACr_2O_4$ spinels[30-32] and has been discussed in relation to $CuCrO_2$ delafossite.[27] The changes in all three lattice parameters when T ~ $T_N$ support a spin driven structural distortion, however further experiments are required to investigate the relationship between the structure and magnetic ordering in α-$SrCr_2O_4$. As is typical for spin-driven structural transitions, the distortions are very small and as such are difficult to detect. More detailed, higher resolution, diffraction measurements around $T_N$ could reveal a lowering of the symmetry to a monoclinic or triclinic unit cell in α-$SrCr_2O_4$. A transition of this kind has been observed in α-$NaMnO_2$, a highly distorted triangular lattice in which a structural transition from monoclinic to triclinic symmetry at $T_N$ results in commensurate magnetic ordering.[23]

The temperature dependence of the lattice parameters and structure in the related compound α-$CaCr_2O_4$ on cooling below room temperature have not been reported, it is thus not possible to make a complete comparison to the properties of α-$SrCr_2O_4$. At room temperature the structures of the two compounds are similar with the most significant difference being the expansion of the unit cell, primarily in the *a* direction, when Ca is replaced by Sr. The relative increase in the *b* and *c* parameters results in a shift in the *c*/*b* ratio, from 0.88 for α-$CaCr_2O_4$[16] to 0.87 α-$SrCr_2O_4$,[14] indicating that the $CrO_2$ layers in α-$CaCr_2O_4$ are more distorted from the ideal hexagonal lattice. Magnetic measurements show very similar behaviour for the two compounds; both order antiferromagnetically, $T_N$ = 43 K with helical magnetic ordering, **k** ~ (0,⅓,0). Differences in the behaviour include the presence of possible lower local symmetry in α-$SrCr_2O_4$ and an increase in the degree of incommensuration of the propagation vector (for α-$CaCr_2O_4$ **k** = (0,0.3317,0)[8]). It is possible that these differences are due to differences in the orthorhombicity of the $CrO_2$ planes in the two compounds, however despite the changes induced by replacing the Ca with the larger Sr cation the magnetic ordering does not appear to differ significantly. Low temperature high resolution electron microscopy or high resolution diffraction may be of interest to resolve the structure of α-$SrCr_2O_4$.

**Acknowledgements**
The authors wish to thank T. M. McQueen for helpful discussions. This research was supported by the US Department of Energy, Division of Basic Energy Sciences, Grant DE-FG02-08ER46544. Use of the National Synchrotron Light Source, Brookhaven National Laboratory, was supported by the U.S.

Department of Energy, Office of Science, Office of Basic Energy Sciences, under Contract No. DE-AC02-98CH10886. Use of the Spallation Neutron Source was supported by the Division of Scientific User Facilities, Office of Basic Energy Sciences, US Department of Energy, under contract DE-AC05-00OR22725 with UT-Battelle, LLC.

Table 1: Structural parameters of α-SrCr$_2$O$_4$ from refinement of powder neutron diffraction data at various temperatures. Space group P*mmn*, Z = 4.

| α-SrCr$_2$O$_4$ | | 100 K | 50 K | 12 K |
|---|---|---|---|---|
| *a* / Å | | 11.62750(11) | 11.62834(10) | 11.63013(9) |
| *b* / Å | | 5.87560(7) | 5.87350(7) | 5.87184(6) |
| *c* / Å | | 5.10292(7) | 5.10125(6) | 5.10019(5) |
| *V* / Å$^3$ | | 348.62(1) | 348.41(1) | 348.29(1) |
| Sr1 2b ¼,¾,*z* | *z* <br> B$_{iso}$ / Å$^2$ | 0.1411(8) <br> 0.76(8) | 0.1407(7) <br> 0.68(7) | 1407(68) <br> 0.65(7) |
| Sr2 2a ¾,¾,*z* | *z* <br> B$_{iso}$ / Å$^2$ | 0.4530(7) <br> 0.35(8) | 0.4525(6) <br> 0.27(7) | 0.4523(5) <br> 0.24(7) |
| Cr1 4f *x*,¼,*z* | *x* <br> *z* <br> B$_{iso}$ / Å$^2$ | 0.5039(4) <br> 0.4993(10) <br> 0.383(11) | 0.5046(4) <br> 0.4981(9) <br> 0.46(9) | 0.5049(4) <br> 0.4975(8) <br> 0.40(9) |
| Cr2 4c 0,0,0 | B$_{iso}$ / Å$^2$ | 0.549(11) | 0.45(9) | 0.37(9) |
| O1 4f *x*,¼,*z* | *x* <br> *z* <br> B$_{iso}$ / Å$^2$ | 0.5909(3) <br> 0.1627(8) <br> 0.22(8) | 0.5912(3) <br> 0.1638(7) <br> 0.26(7) | 0.5910(2) <br> 0.1649(6) <br> 0.23(7) |
| O2 4f *x*,¼,*z* | *x* <br> *z* <br> B$_{iso}$ / Å$^2$ | 0.4152(3) <br> 0.8231(8) <br> 0.29(9) | 0.4148(2) <br> 0.8226(7) <br> 0.34(8) | 0.4143(2) <br> 0.8233(6) <br> 0.22(7) |
| O3 8g *x,y,z* | *x* <br> *y* <br> *z* <br> B$_{iso}$ / Å$^2$ | 0.40605(15) <br> 0.4997(4) <br> 0.3333(6) <br> 0.23(6) | 0.40612(13) <br> 0.4999(4) <br> 0.3333(6) <br> 0.26(6) | 0.40617(12) <br> 0.4993(4) <br> 0.3337(5) <br> 0.21(5) |

Table 2: Bond Lengths in α-SrCr$_2$O$_4$ at various temperatures obtained from powder neutron diffraction.

| α-SrCr$_2$O$_4$ | 100 K | 50 K | 12 K |
|---|---|---|---|
| Sr1-Sr2 / Å | 3.595(3)x2<br>4.222(4)x2 | 3.596(2)x2<br>4.216(3)x2 | 3.596(2)x2<br>4.215(3)x2 |
| Sr1-O1 / Å | 2.414(5)x2 | 2.413(4)x2 | 2.418(3)x2 |
| Sr1-O3 / Å | 2.533(3)x4 | 2.534(3)x4 | 2.537(2)x4 |
| <Sr1-O> / Å | 2.49 | 2.49 | 2.50 |
| Sr2-O2 / Å | 2.382(4)x2 | 2.376(3)x2 | 2.372(3)x2 |
| Sr2-O3 / Å | 2.576(3)x4 | 2.578(3)x4 | 2.575(2)x4 |
| <Sr2-O> / Å | 2.51 | 2.51 | 2.51 |
| Cr1-O1 / Å | 1.993(6) | 1.980(6) | 1.970(5) |
| Cr1-O2 / Å | 1.948(6) | 1.957(6) | 1.967(5) |
| Cr1-O3 / Å | 1.997(4)x2<br>2.041(4)x2 | 1.984(4)x2<br>2.043(4)x2 | 1.995(4)x2<br>2.039(4)x2 |
| <Cr1-O> / Å | 2.00 | 2.00 | 2.00 |
| Cr2-O1 / Å | 1.991(3)x2 | 1.995(2)x2 | 1.996(2)x2 |
| Cr2-O2 / Å | 1.986(3)x2 | 1.989(2)x2 | 1.990(2)x2 |
| Cr2-O3 / Å | 2.021(3)x2 | 2.021(2)x2 | 2.022(2)x2 |
| <Cr2-O> / Å | 2.00 | 2.00 | 2.00 |
| Cr1-Cr1 (in-plane) / Å | 2.939(0)x2 | 2.939(0)x2 | 2.938(0)x2 |
| Cr2-Cr2 (in-plane) / Å | 2.938(0)x2 | 2.937(0)x2 | 2.936(0)x2 |
| Cr1-Cr2 (in-plane) / Å | 2.941(4)x2<br>2.948(4)x2 | 2.935(4)x2<br>2.952(4)x2 | 2.932(4)x2<br>2.954(4)x2 |
| <Cr-Cr> (in-plane) / Å | 2.94 | 2.94 | 2.94 |
| Cr1-Cr1 (inter-layer) / Å | 5.701(5)<br>5.929(5) | 5.707(7)<br>5.921(7) | 5.723(7)<br>5.904(7) |
| Cr2-Cr2 (inter-layer) / Å | 5.815(0)x2 | 5.814(0)x2 | 5.814(0)x2 |
| <Cr-Cr> (inter-layer) / Å | 5.82 | 5.81 | 5.81 |

**Figure Captions**

Figure 1: (a) Structural model of α-SrCr$_2$O$_4$ (b) the projection of the CrO$_2$ layers onto the *bc*-plane, the Cr-Cr distances are those obtained from neutron powder diffraction at 50 K and show the very small distortion from an ideal hexagonal lattice. CrO$_6$ octahedra for the crystallographically distinct Cr1 and Cr2 sites are shown in blue and orange respectively. The Sr$^{2+}$ cations, situated between the CrO$_2$ layers are light grey.

Figure 2: Magnetic susceptibility (M/H) as a function of temperature for α-SrCr$_2$O$_4$. The inverse susceptibility (H/M) is inset.

Figure 3: Specific heat of α-SrCr$_2$O$_4$ as a function of temperature. An estimate of the change in magnetic entropy using the Debye model to subtract the lattice entropy, $T < 75$ K, is inset.

Figure 4: Temperature dependence of the lattice parameters and volume of α-SrCr$_2$O$_4$ as determined from synchrotron X-ray diffraction. For an ideal hexagonal *bc*-plane $b\cos30 = c$. The solid lines are a guide to the eye, unless shown the error bars are smaller than the data points.

Figure 5: Observed (red points) and calculated (black line) neutron diffraction data for α-SrCr$_2$O$_4$ collected at 50 K. The difference curve is also shown; reflection positions for the main phase of α-SrCr$_2$O$_4$ (upper) and a Cr$_2$O$_3$ impurity (0.33(1) wt%) (lower) are indicated by the vertical lines.

Figure 6: (a) Observed (red points) and calculated (black line) neutron diffraction data for α-SrCr$_2$O$_4$ collected at 12 K. The difference curve is also shown; reflection positions for the nuclear (upper) and magnetic (middle) phases of α-SrCr$_2$O$_4$ and a Cr$_2$O$_3$ impurity (0.29(2) wt%) (lower) are indicated by the vertical lines. Fits to the magnetic peak at d ~ 4.15 Å for **k** = (0,0.3217(8),0) (lower) and **k** = (0,⅓,0) (upper) are shown in more detail in the inset. (b) The magnetic structure of α-SrCr$_2$O$_4$. Cr1 and Cr2 atomic sites are shown in blue and orange respectively. (c) The evolution of the magnetic peaks at d ~ 4.15 Å as a function of temperature.

Figure 1

(a) 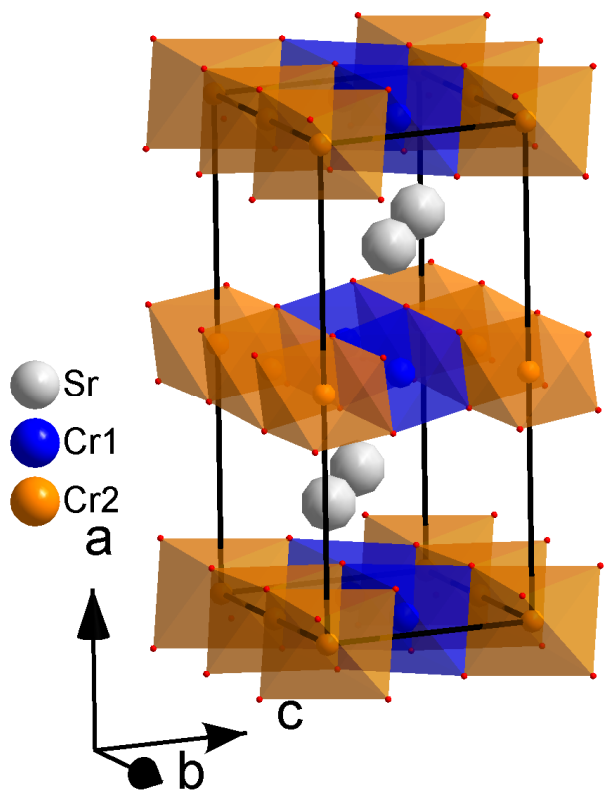

(b) 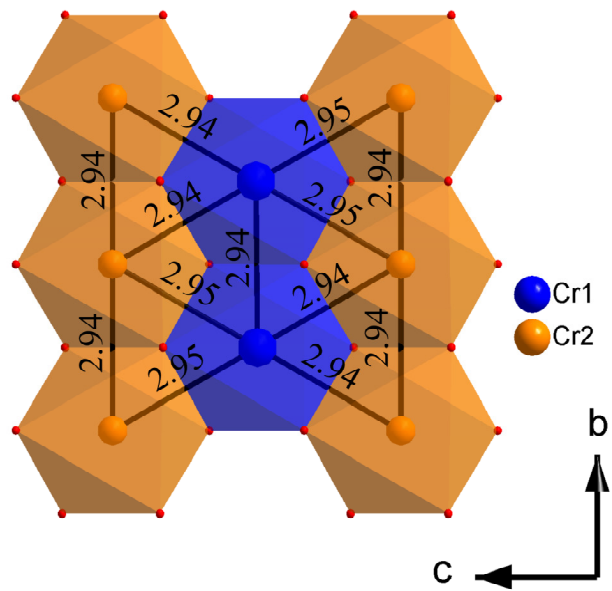

Figure 2

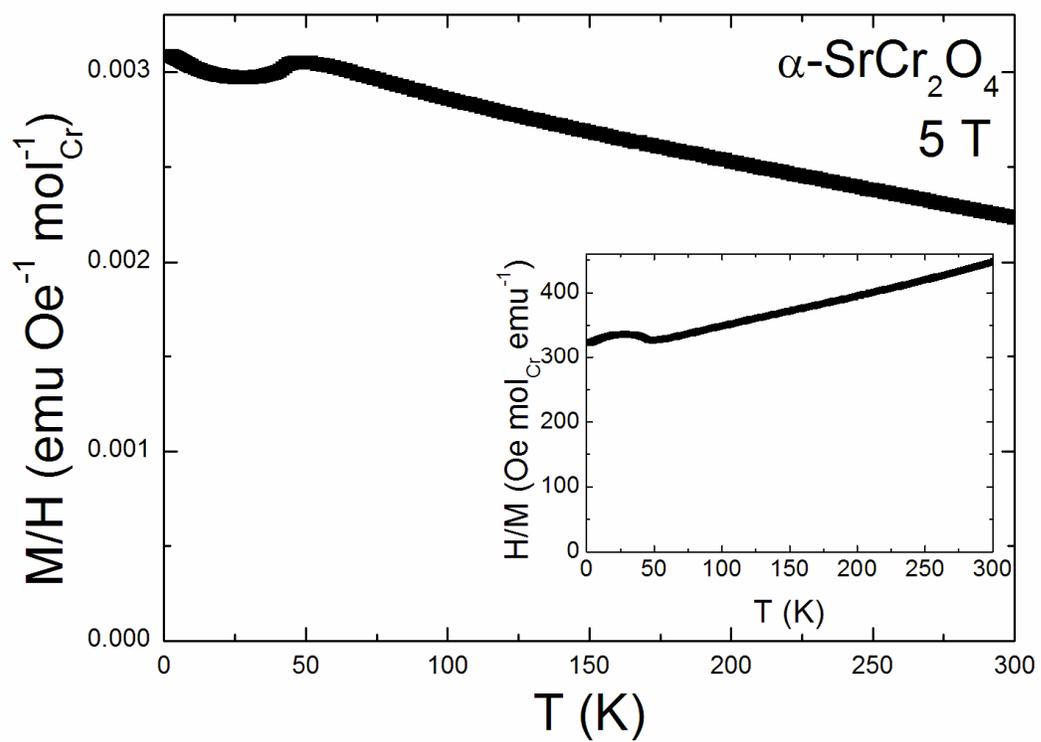

Figure 3

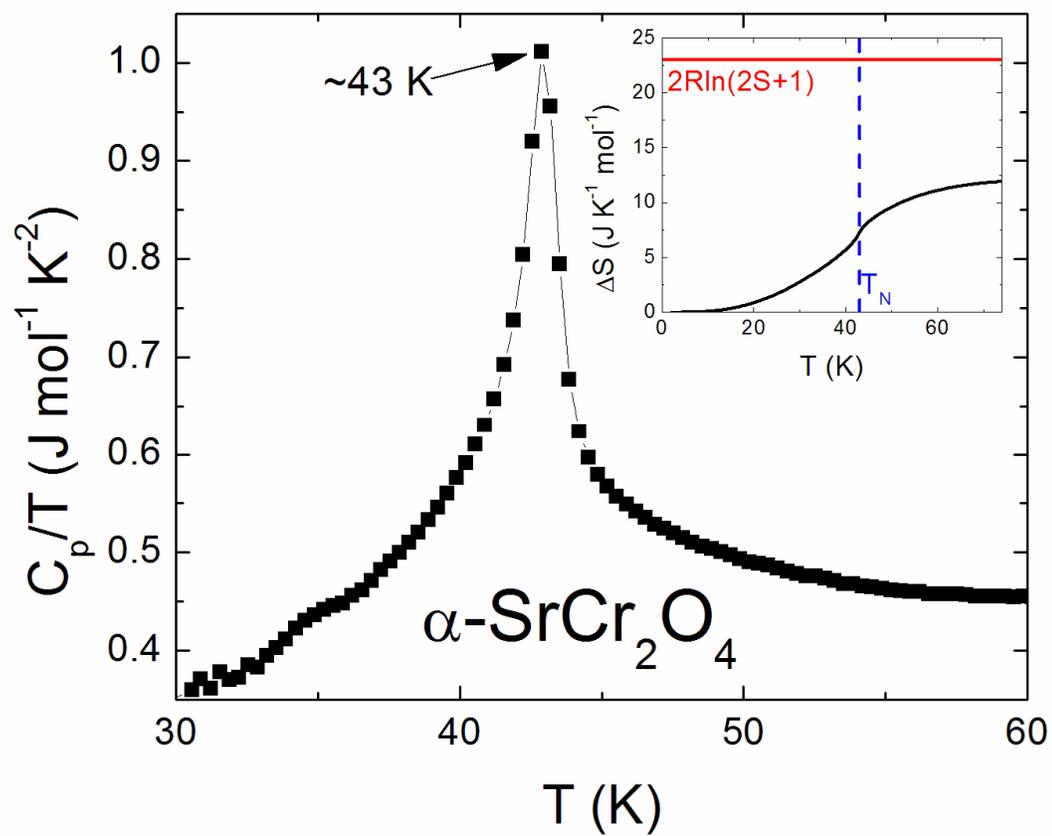

Figure 4

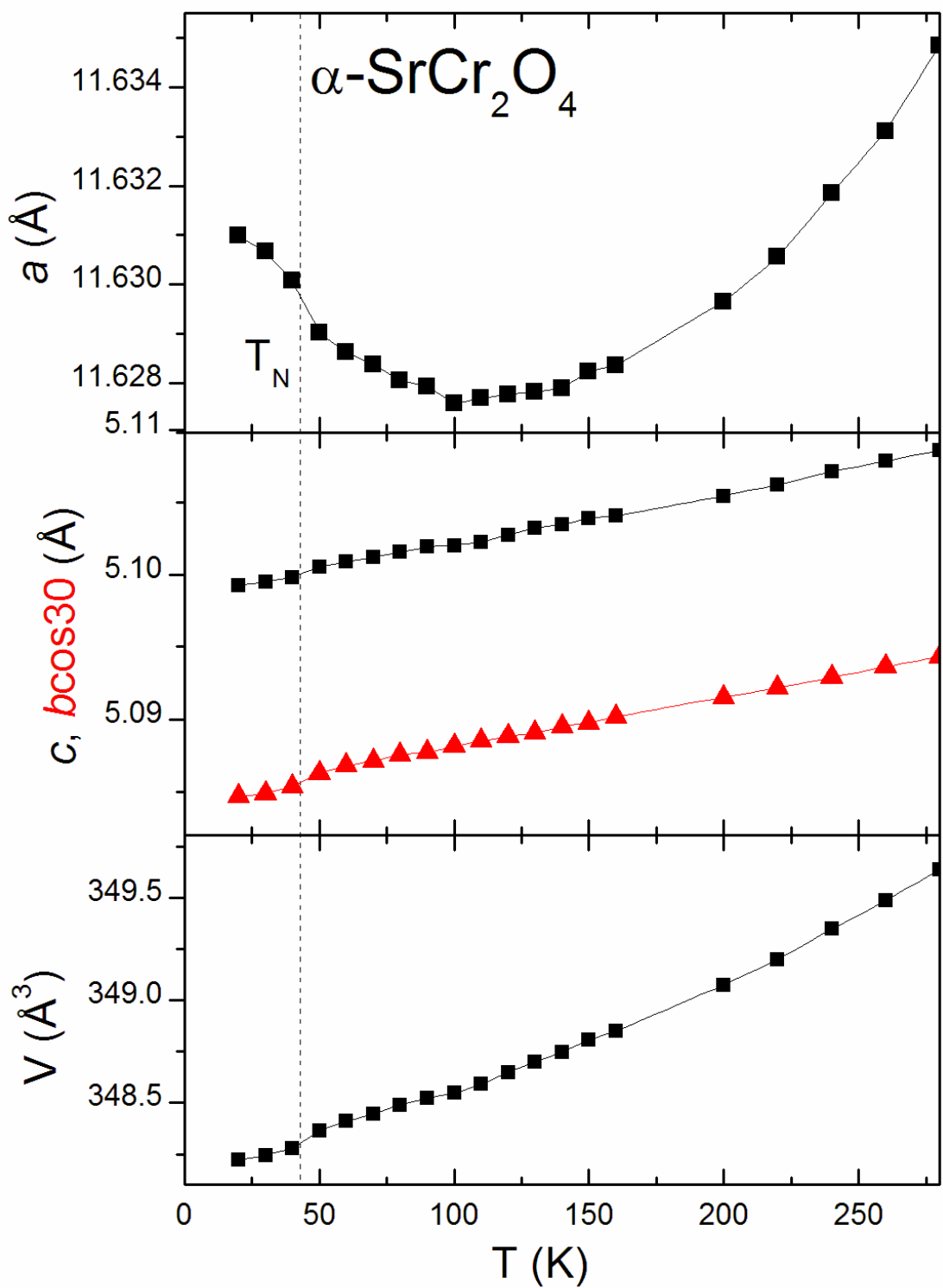

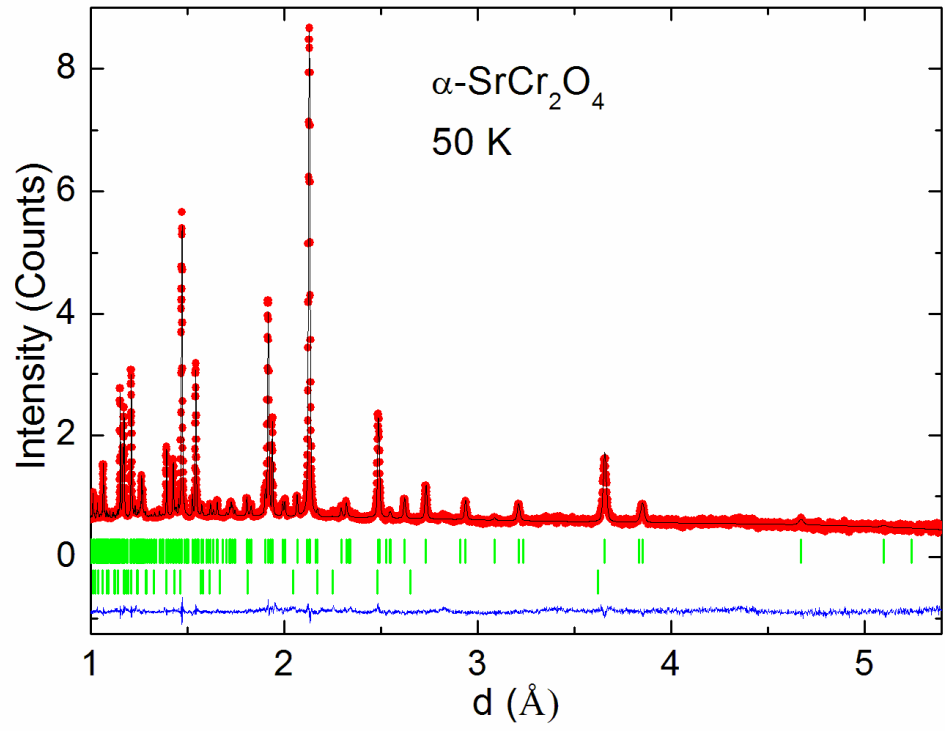

Figure 6

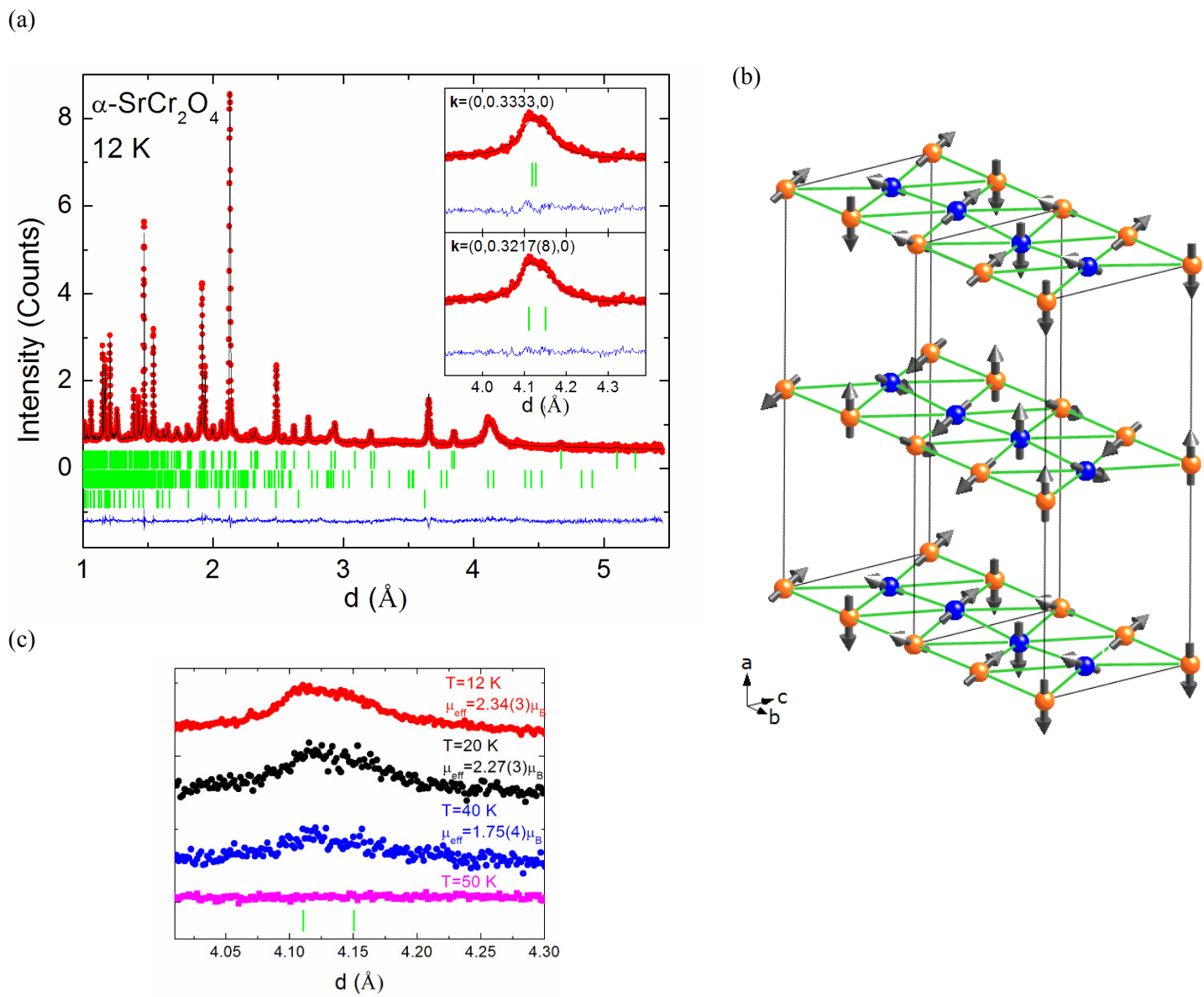